\documentclass[letterpaper, 10 pt, conference]{ieeeconf}

\IEEEoverridecommandlockouts 
\usepackage[USenglish]{babel} 
\usepackage[T1]{fontenc}
\usepackage[ansinew]{inputenc}
\usepackage[noadjust]{cite}
\usepackage{lmodern} 
\usepackage{graphicx} 
\usepackage{float}

\usepackage{amsmath}
\usepackage{amsfonts}
\usepackage{bm}
\usepackage{url}

\newcommand{\dsr}{d_{\text{sr}}}
\newcommand{\dpyl}{d_{\text{pyl}}}
\newcommand{\droad}{d_{\text{road}}}

\newcommand{\m}{\text{m}}
\newcommand{\s}{\text{s}}
\newcommand{\ymin}{y_{\text{min}}}
\newcommand{\ymax}{y_{\text{max}}}

\begin{document}


\title{Designing a Roadside Sensor Infrastructure to Support Automated Driving}

\author{
\authorblockN{Florian Geissler}\\
\authorblockA{Dependability Research Lab, 
Intel Labs Europe\\
85579 Neubiberg, Germany\\
Email: florian.geissler@intel.com}
\and
\authorblockN{S{\"o}ren Kohnert and Reinhard Stolle}\\
\authorblockA{Augsburg University of Applied Sciences \\
86161 Augsburg, Germany \\
Email: soeren.kohnert@hs-augsburg.de}
}

\date{\today}
\maketitle



\begin{abstract}
Automation of complex traffic scenarios is expected to rely on input from a roadside infrastructure to complement the vehicles' environment perception.
We here explore design requirements for a prototypical setup of virtual vision or RADAR sensors along one roadside. 
Explicitly, we analyze the road coverage and the probability of vehicle occlusions, with the objective of evaluating the completeness of information that is captured by the sensor field.
Simulation case studies are performed based on real traffic data acquired at the German Autobahn 9 near Munich.
Our findings indicate how the sensor network should be designed in terms of sensor range, orientation and opening angle, in order to enable effective traffic detection.
The achieved degree of completeness suggests that such a setup could be used to support automated vehicles to a substantial extent.
\end{abstract}

\section{Introduction}

For the last decades, the realization of automated driving (AD) has experienced tremendous progress, powered by continuous technological innovations throughout a variety of fields. Improved object detection techniques (see e.g. \cite{Sun2006, Geiger2012}), data fusion and tracking schemes \cite{Kalman1960, Danescu2011a}, dedicated wireless communication paths \cite{DSRC1999}, as well as smart driving strategies \cite{Shalev-Shwartz2017}, have contributed to the functional safety and dependability of AD.
So far, most concepts focus on an in-vehicle perspective: Automated cars are equipped with various sensors, comprising for instance camera, RADAR and LIDAR, to be able to detect threats in their environment. 
It appears, however, that this development faces some important challenges. First, the vehicle horizon is limited to the range of the sensors on board. Second, the processing of data collected by all sensors in the vehicle demands a lot of computational resources.
To overcome such difficulties, one approach is to share information and resources with other agents on the road, or with the infrastructure. Corresponding technologies are commonly referred to as vehicle-to-vehicle (V2V) and vehicle-to-infrastructure (V2I), respectively (e.g. \cite{Wu2004, Santa2008}). The latter is typically understood as the exchange of information with individual elements of an existing infrastructure, such as traffic lights.

In this paper, we extend this concept to study a system where the perception of road traffic, and therefore the information relevant for an autonomous agent, is exclusively acquired by the infrastructure. This not only shifts the computational load off the vehicle, but also provides long-range vision for early road hazard warnings and traffic flow management. The infrastructure is in this case represented by an array of sensors mounted along one roadside at fixed spatial intervals. 
This particular setup is motivated by a highway scenario where a sensor network for road surveillance can be established, exploiting existing equipment, by attaching each sensor to one of the roadside guiding posts. 
The fact that we restrict ourselves to a single roadside here reflects the practical difficulties of placing sensors on the median strip of the highway.
Detections of individual sensors are collected, processed in an close-by edge computation platform, and are eventually broadcast to the vehicle. 
We particularly focus on RADAR sensors here, even though the first part of our analysis applies to any generic type of sensor, that is specified only by the form of its field of view (FoV).
RADAR sensors were shown to be feasible instruments for vehicle detection \cite{Fang2007, Felguera-Martin2012, Dickmann2016}, however, to the best of our knowledge, the possibility of a seamless road surveillance by a network of RADAR sensors has not been studied so far. 

A crucial concern for the type of setup described above, is whether the entirety of information obtained by the infrastructure is of an adequate quality to dictate safe driving decisions. In this paper, we address this question with respect to the completeness of information, which is one of the paramount quality attributes in this context (see e.g. \cite{Lee2002}), and a prerequisite for the system's functional safety. Hereby, we define completeness as the extent to which data are of sufficient scope to represent the universe of discourse \cite{Wang1996, Wiltschko2004}. 
We thus analyze, to what degree the system captures objects on the road, simulating different detection mechanisms such as vision and RADAR sensing, and using realistic road data. 
Completeness of information is generally challenged by false negative detections, that can occur for the following reasons:
i) An object is missed because it is not in the sensors FoV, ii) an object is in the sensors FoV, but the infrastructure fails to detect it, for instance due to a low signal-to-noise ratio (SNR), iii) an object can not be detected because it is occluded by other objects.
False positive detections further complicate traffic detection, however, are typically far less safety-critical than false negatives \cite{Shalev-Shwartz2017}. They do not affect the information completeness, and are therefore not studied explicitly here.
Importantly, the specifications of the setup differ from previous work (see e.g.\cite{Cheng2010, Dash2018}). A special emphasis is put on the influence of vehicle occlusions on a seamless road surveillance. The relation between occlusions and traffic density, based on sample measurements, was for instance explored in \cite{Schopplein2013}. The goal of the present paper is to investigate the fitness of a roadside sensor infrastructure for the purpose of guiding automated vehicles.
Our analysis can be used to design sensor network topologies with a high degree of detection completeness. 

The article is organized as follows: In Sec.~\ref{sec:roadCoverage} we analytically derive the requirements for full and multiple road coverage in the given setup, in the absence of road traffic. 
In order to avoid systematic incompleteness of traffic detection, full coverage is a key aspect for the design of a dependable sensor network. 
Subsequently, in Sec.~\ref{sec:occlusions}, we address the problem of detecting moving objects taking into account vehicle occlusions.
We discuss several simulation results that are based on different detection mechanisms and traffic conditions. Finally, we conclude in Sec.~\ref{sec:summary}.

\section{Road coverage}
\label{sec:roadCoverage}

\subsection{Setup}

\begin{figure}[h]
\center
\includegraphics[width=1\linewidth]{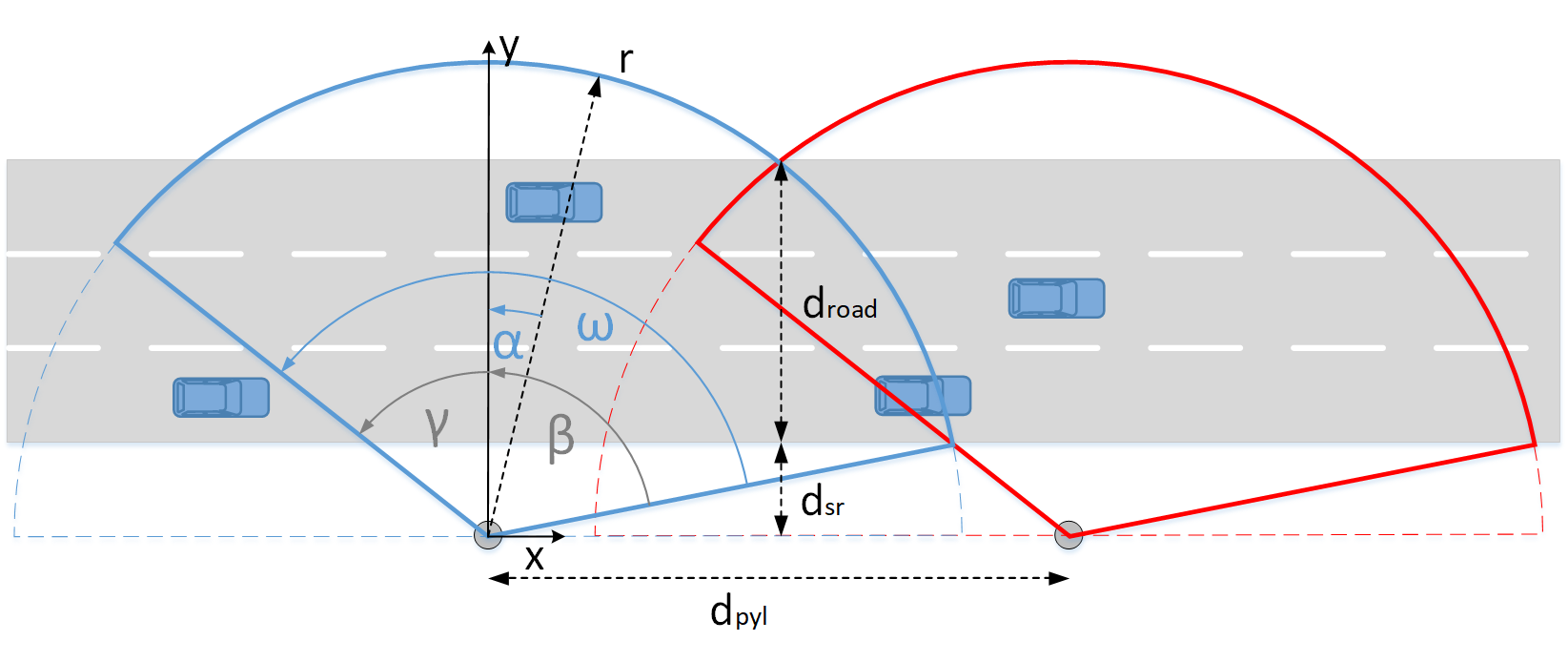}
\caption{Sketch of the sensor setup, and the relevant tunable parameters. Due to a periodic arrangement of the sensors, it is sufficient to consider two neighboring sensors only. Positive angles are as usual defined in anti-clockwise orientation.}
\label{fig:Setup}
\end{figure}

In the following, we consider a generic sensor model of reduced complexity.
All sensors forming the network are identical, oriented the same way, and arranged periodically along a single, straight roadside. 
The symmetries of the system allow to study the overlap of two neighboring sensor sectors only, in order to analyze the requirements for full road coverage of the entire sensor network. In this section, we derive the positions of the two possible intersection points of a pair of neighboring sensor FoVs. Full road coverage is given if the road boundaries are within the corridor specified by those lower and upper intersection points.
Furthermore, we presume that a sensor FoV is two-dimensional and has the geometric form of a circle sector, thus it can be parametrized by opening angle and range. 
The defining parameters of the sensor array topology are illustrated in Fig.~\ref{fig:Setup}: sensor range ($r$), sensor opening angle ($\omega$), rotation angle of a sensor relative to the $y$-axis ($\alpha$), distance of a sensor to the nearest road boundary ($\dsr$), distance between two neighboring sensors, i.e. two neighboring pylons, ($\dpyl$), and road width ($\droad$).
For convenience, let us also define the angles $\beta=\omega/2+\alpha$ and $\gamma=\omega/2-\alpha$.

A sensor FoV that points away from the road does not yield any additional information, and the structure can always be transformed into a more efficient one by modifying $\omega$ and $\alpha$ accordingly. Therefore, we restrict ourselves to
\begin{align}
& 0 \leq \omega \leq \pi, \hspace{2cm} |\alpha|\leq \frac{1}{2}(\pi -\omega).
\label{eq:w0}
\end{align}

\subsection{Requirements for full road coverage}
\label{sec:reqCov}

To derive the requirements for full coverage, we have to find the lower and upper intersection points of two overlapping sensor sectors.
The former can emerge either as the intersection of two lines representing the lateral edges of the FoV, or as the intersection of one boundary line with a circle arc, representing the front edge of a sensor FoV. Similarly, the latter intersection point can arise as the intersection of a line and a circle arc, or as the intersection of two circle arcs, leaving us with a total of four different geometrical constellations. 

We first study the lower intersection point. The lines describing the left and right lateral edge of a sensors FoV are given by
\begin{align}
x_r= y_r\tan(\beta), \hspace{2cm}
x_l=-y_l \tan(\gamma).
\label{eq:defXy}
\end{align}
Here, $x_r$ ($x_l$) is the lateral distance of a point on the right (left) boundary line from its original sensor, while the vertical distances are given by $y_r$ and $y_l$, respectively. The boundary lines of two neighboring sensors cross at
\begin{equation}
x_r=\dpyl+x_l, \hspace{2cm} y_r=y_l=y.
\label{eq:defCrossing}
\end{equation}
With (\ref{eq:defXy}), this leads us to
\begin{align}
\ymin=y=\frac{\dpyl}{\tan(\beta)+\tan(\gamma)}.
\end{align}
This defines the lower intersection point in case the sensor range is sufficiently large, so if the crossing point is in reach of both sensors,
$r\geq \max \left(\ymin/\cos(\beta), \ymin/\cos(\gamma) \right)= \ymin/\cos(\omega/2+|\alpha|)$.
It follows that the first requirement for full road coverage can be phrased as
\begin{align}
\ymin \leq \dsr,
\hspace{1cm} \text{if}\ r \geq 
\frac{\ymin}{\cos(\omega/2+|\alpha|)}.
\label{eq:req1a}
\end{align}

Second, if the sensor range is not sufficient for the above requirement, the sensor sectors can still overlap if the crossing point emerges as an intersection of one line and one circle arc. We have
\begin{align}
x_{r}=\sqrt{r^2-y^2}, \hspace{2cm}
x_{l}=-y \tan \left(\frac{\omega}{2}-|\alpha|\right),
\end{align}
where we use $\min(\beta,\gamma)=\omega/2-|\alpha|$ to account for both cases of $\alpha>0$ and $\alpha<0$.
The constraints for the crossing point in (\ref{eq:defCrossing}) lead us to 
\begin{align}
\sqrt{r^2-y^2}+ y \tan \left(\frac{\omega}{2}-|\alpha|\right) =\dpyl,
\end{align}
with the two solutions
\begin{align}
y_{\pm}
&= \cos \left(\frac{\omega}{2}-|\alpha|\right) \notag \\
&\times \left( \dpyl \sin\left(\frac{\omega}{2}-|\alpha|\right) \pm \sqrt{r^2-\dpyl^2\cos^2\left(\frac{\omega}{2}-|\alpha|\right)}\right).
\label{eq:ypm2a}
\end{align}
Those solutions are real (an intersection exists) if $r\geq \dpyl\cos(\omega/2-|\alpha|)$.
To identify the lower crossing point in this case, we choose the smaller solution $y_-$ (lower sign) and therefore obtain
the requirement for full coverage,

\begin{align}
\label{eq:req1b}
& y_- \leq \dsr,  \\
& \text{if}\ r<\frac{\ymin}{\cos(\omega/2+|\alpha|)}, \frac{\omega}{2}+|\alpha|\geq \arccos\left(\frac{y_-}{r}\right), \notag\\
& \text{and}\ r\geq \max\left(\frac{y_-}{\cos(\omega/2-|\alpha|)}, \dpyl\cos\left(\frac{\omega}{2}-|\alpha|\right) \right). \notag 
\end{align}
Here, an additional constraint needed to be taken into account -- the range has to be sufficiently large to reach the crossing point at all, 
$r \geq y_{-}/\cos(\omega/2-|\alpha|)$ and $\max(\beta,\gamma)=\omega/2+|\alpha| \geq \arccos(y_{-}/r)$.
These terms were added in the last two lines of (\ref{eq:req1b}).

Next we study the vertical distance of the upper crossing point. As before, we have to distinguish two cases.
First, the intersection can emerge as the crossing point of two circle arcs. This is the case if the opening angle is wide enough, explicitly,
\begin{align}
&\frac{\omega}{2}-|\alpha| \geq \frac{\pi}{2}- \arcsin\left( \frac{\dpyl}{2r}\right).
\end{align}
In this case, the crossing point is simply given by 
\begin{align}
y=\ymax=\sqrt{r^2-\left(\frac{\dpyl}{2}\right)^2},
\label{eq:defYmax}
\end{align}
which brings us to the requirement for full coverage of 
\begin{align}
\label{eq:req2a}
&\ymax\geq \dsr+\droad, \\
&\text{if}\ \frac{\omega}{2}-|\alpha| \geq \frac{\pi}{2}- \arcsin\left(\frac{\dpyl}{2r}\right)\ \text{and}\ r\geq \frac{\dpyl}{2}. \notag
\end{align}

Furthermore, the upper crossing point can arise as an intersection point of a circle arc and a line. We then make use of (\ref{eq:ypm2a}), where the desired maximum distance is now associated with the upper sign, $y_+$, and full road coverage is given if
\begin{align}
\label{eq:req2b}
& y_+ \geq \dsr+\droad, \\
&\text{if}\ \frac{\omega}{2}-|\alpha|< \frac{\pi}{2}- \arcsin\left(\frac{\dpyl}{2r}\right),\ \frac{\omega}{2}+|\alpha| \geq \arccos\left( \frac{y_{+}}{r}\right), \notag \\
& \text{and}\ r\geq  \max\left(\frac{y_{+}}{\cos(\omega/2-|\alpha|)}, \dpyl \cos\left(\frac{\omega}{2}-|\alpha|\right) \right). \notag
\end{align}
As before, we check for sufficient sensor range by
$r \geq y_{+}/\cos(\omega/2-|\alpha|)$ and $\omega/2+|\alpha| \geq \arccos(y_{+}/r)$.

Full road coverage is achieved if (\ref{eq:req1a}) or (\ref{eq:req1b}), and either of (\ref{eq:req2a}) and (\ref{eq:req2b}) holds.

\subsection{Implications}

The findings of the above section allow us to derive analytically the threshold between the regimes of incomplete and complete road coverage.
We note the following implications. 

The minimum sensor range that possibly provides full road coverage is determined by the points the furthest away from any sensor. This distance is given by (\ref{eq:defYmax}), and we find
\begin{equation}
r_{\text{min}}= \sqrt{\left(\frac{\dpyl}{2}\right)^2 + \left(\dsr+\droad \right)^2}.
\end{equation}
The angle $\omega$ on the other hand has no lower bound -- for infinitely large range it approaches zero.

\begin{figure}[ht]
\centering
  \includegraphics[width=1\linewidth]{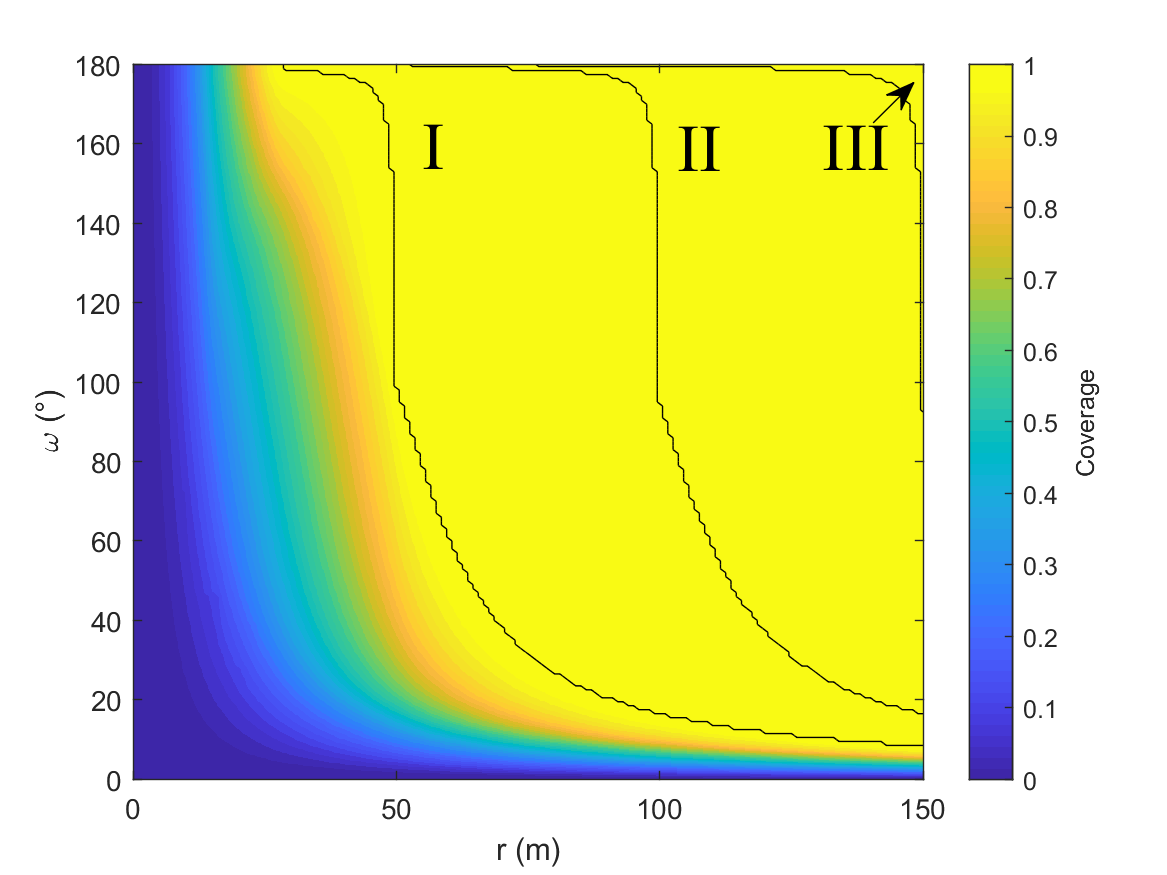}
\caption{Calculation of the road coverage by numerical integration (contour plot), and the exact threshold boundaries for full coverage as derived with (\ref{eq:req1a}), (\ref{eq:req1b}), (\ref{eq:req2a}), and (\ref{eq:req2b}), for single (I), double (II), and triple (III) coverage. Parameters in this example were chosen to be $\dpyl=50\m$, $\droad=14\m$, $\dsr=0.5\m$, and $\alpha =(\pi -\omega)/2$.}
  \label{fig:WRcoverage}
\end{figure}

Without a repositioning of the sensors, road coverage can be optimized by varying dynamically the parameters $r$, $\omega$ and $\alpha$.
Mounting the sensors very close to the road, such that in particular $\dsr \ll \dpyl$ (which seems realistic for our scenario), is a challenge to full road coverage. A large rotation angle $\alpha$, almost to the full extent of $\alpha \approx (\pi -\omega)/2$, is required then. 
We therefore fix $\alpha$ in the following to this maximum, and investigate the impact of the remaining parameters $\omega$ and $r$ on the road coverage.
Fig.~\ref{fig:WRcoverage} gives an example, and illustrates the validity of the above analysis by comparing it to the outcome of a numerical integration approach.

Importantly, the analysis of Sec.~\ref{sec:reqCov} can be easily extended to identify regimes of multiple road coverage. To do so, we replace 
in (\ref{eq:req1a}), (\ref{eq:req1b}), (\ref{eq:req2a}), and (\ref{eq:req2b}) the parameter $\dpyl$ by $n\cdot \dpyl$, where $n$ is a positive integer.
We then obtain the constraints for full road coverage by a subset of sensors that are $n$-nearest neighbors. This is equivalent to the statement that each point of the road is in range of at least $n$ sensors, so we face multiple road coverage of degree $n$ (see also Fig.~\ref{fig:WRcoverage}). 
This finding can be used to design a robust sensor network that maintains full road coverage despite temporary sensor outages.

By inspection of Fig.~\ref{fig:WRcoverage}, we state that full (and even multiple) road coverage is readily possible with current state-of-the-art RADAR technology \cite{Dickmann2016}.


\section{Occlusions}
\label{sec:occlusions}

\subsection{Traffic model}

In this section, we investigate the influence of object occlusions on the detection capabilities of the sensor network, depending on the sensor FoV.
To account for a realistic traffic situation, we use statistical road traffic data recorded at the Autobahn 9 close to Neufahrn, near Munich, which was provided by the traffic authority of Southern Bavaria \cite{ABDSB2018}. This data contains for instance the average number of vehicles per hour per lane, average speed per lane, and the distribution of vehicle types per lane. The latter follows a classification scheme with five different groups, including cars, trucks, buses, as well as cars and trucks with trailers.
The considered highway segment offers three lanes, with an additional emergency lane that can be opened for vehicles on demand by the traffic control authority. 
We compare three selected scenarios here: Christmas Eve morning, which represents a low-density traffic scenario, an average Tuesday morning, representing typical commuting traffic, and a traffic jam, as an example for a high-density traffic situation. In our simulation we adopt a simplified traffic model. 
The road is randomly populated with vehicles, weighted according to traffic density and object type distributions.
All vehicles on a lane move with the same, constant speed  -- lane changes and related maneuvers are neglected here.

\subsection{Vision sensing}

In the first stage of the simulation, we study occlusions using a line-of-sight (LoS) sensing model.
A ray casting algorithm calculates the shadows thrown by the objects. Explicitly, if at least one resolution cell of a very small size of $5\times 5\text{cm}^2$ of a vehicle's bounding box is illuminated, the car is classified as visible. 
With such LoS sensors we model cameras with almost perfect detection capabilities, that act as a reference detection frame.
The simulation returns the percentage of missed vehicles over time, which equals the degree of incompleteness in this setup.

\subsection{RADAR sensing}

Next, we simulate a network of RADAR sensors using existing MATLAB$^{\textregistered}$ toolboxes \cite{Matlab2017}. 
This provides us with generic RADAR detection generators, that can be used to model object detections without further specification of the signal processing inside a sensor.
The radiation pattern of the RADAR sensors is idealized as conoidal and unambiguous, meaning in particular that sidelobes are neglected. We assume an azimuth angle resolution of $\omega/12$, as it was achieved with a linear array of 12 antennas, where the number of antennas is constant in contrast to the aperture size. Furthermore, we use a radial resolution of $1\m$, and the false alarm rate is quantified as $10^{-6}$ \cite{Richards2014}. 
As before, we assume that an area within the FoV of a given sensor is occluded for RADAR detections if it is shadowed by an object such that there exists no clear LoS to that sensor. 
The RADAR cross section (RCS) of a vehicle is taken as a constant across the object dimensions here for simplicity.
In order to determine the number of missed objects at an instant of time, each vehicle is assigned a unique ID, which is registered for an individual sensor detection. Completeness is then evaluated by comparing the list of detected IDs with the ground truth. Eventually, the time average over the course of the simulation is performed.

We further advance our analysis by studying object tracking based on the RADAR detections. Generally, tracking allows to compensate temporary losses of information in situations where objects are occluded, or out of sensor range. On the other hand, the sensitivity to errors, and the temporal delay of a measurement-to-track association reduces the success rate of object detections. In the following we therefore observe a competition of these two opposing trends.
Before tracks are created, all individual detections are collected by a central authority -- for instance the edge computing platform -- and fused to clusters according to spatial proximity. 
Such a centralized fusion scheme facilitates tracking in situations where an object travels from one sensor sector to another.
As a cluster length, we hereby choose the width of a car, i.e. the smallest object length scale studied in the present setup. Subsequently, clusters are assigned to tracks using a linear Kalman filter (KF), that is based on a two-dimensional constant-velocity motion model. 
A KF uses a series of consecutive measurements to predict the next vehicle state, and updates this estimation by comparing it to the subsequent measurements \cite{Kalman1960}. With the help of such periodic prediction-update cycles, the filter is able to track vehicle trajectories. Key tracking parameters of our model, such as the track confirmation threshold and the number of coasting updates, were set to the standard values specified in \cite{Matlab2017}.

In the simulation, the following aspects further impede a complete track object detection: i) If an object enters the sensor FoV, a track is confirmed only after a threshold number of detections that match with the KF predictions. In the meantime, it is missed. ii) Multiple objects can be assigned to the same track, this applies in particular to two objects in close vicinity on different lanes. iii) False alarm detections due to noise disarray tentative tracks. 
Note that long objects, like trucks and buses in this model, are typically assigned to several tracks at once, however, such redundant false positive detections will not affect completeness here, as discussed above.

\subsection{Results}

\begin{figure}[tpb]
\centering
\includegraphics[width=1\linewidth]{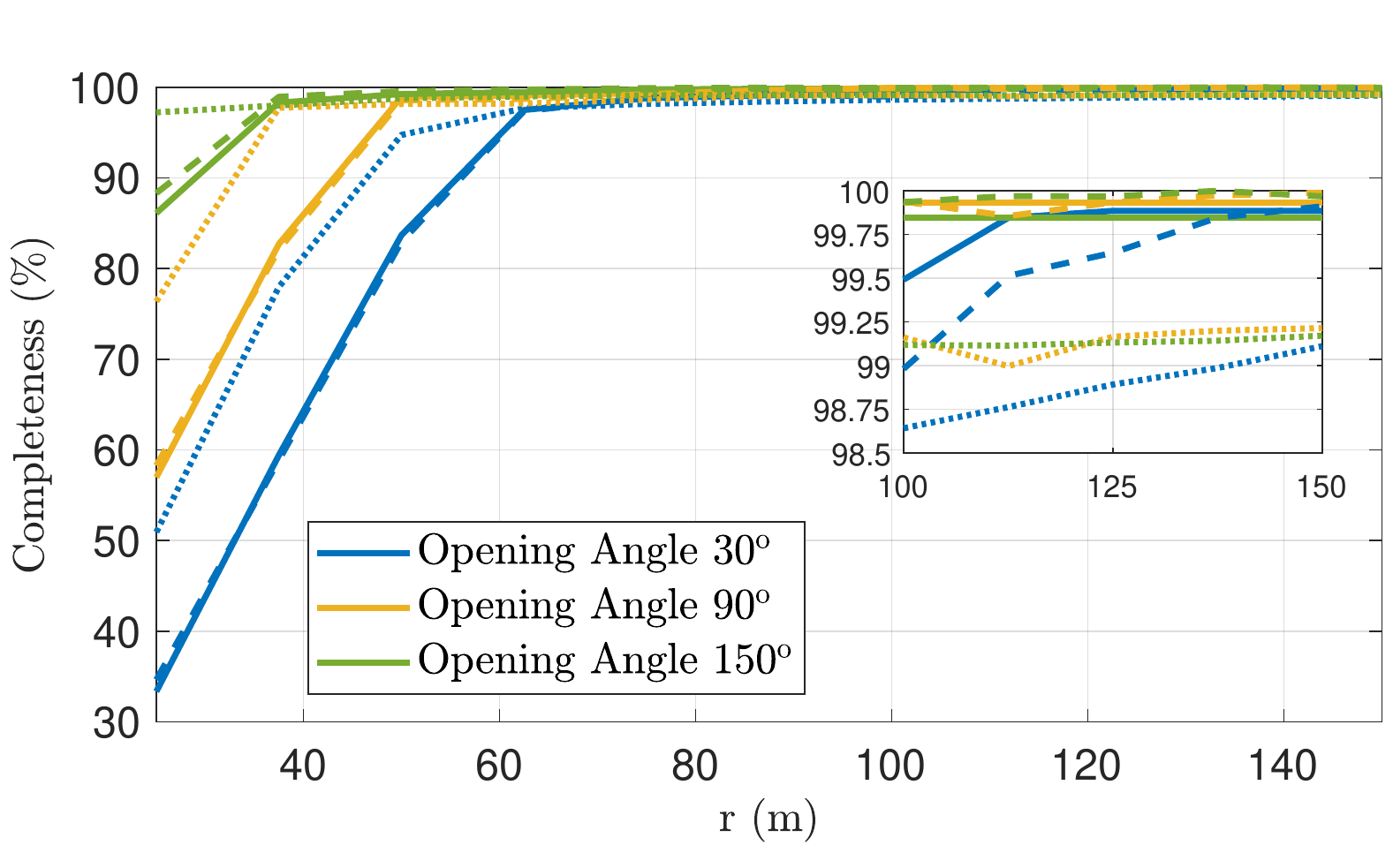}\\
a) Christmas Eve\\
\includegraphics[width=1\linewidth]{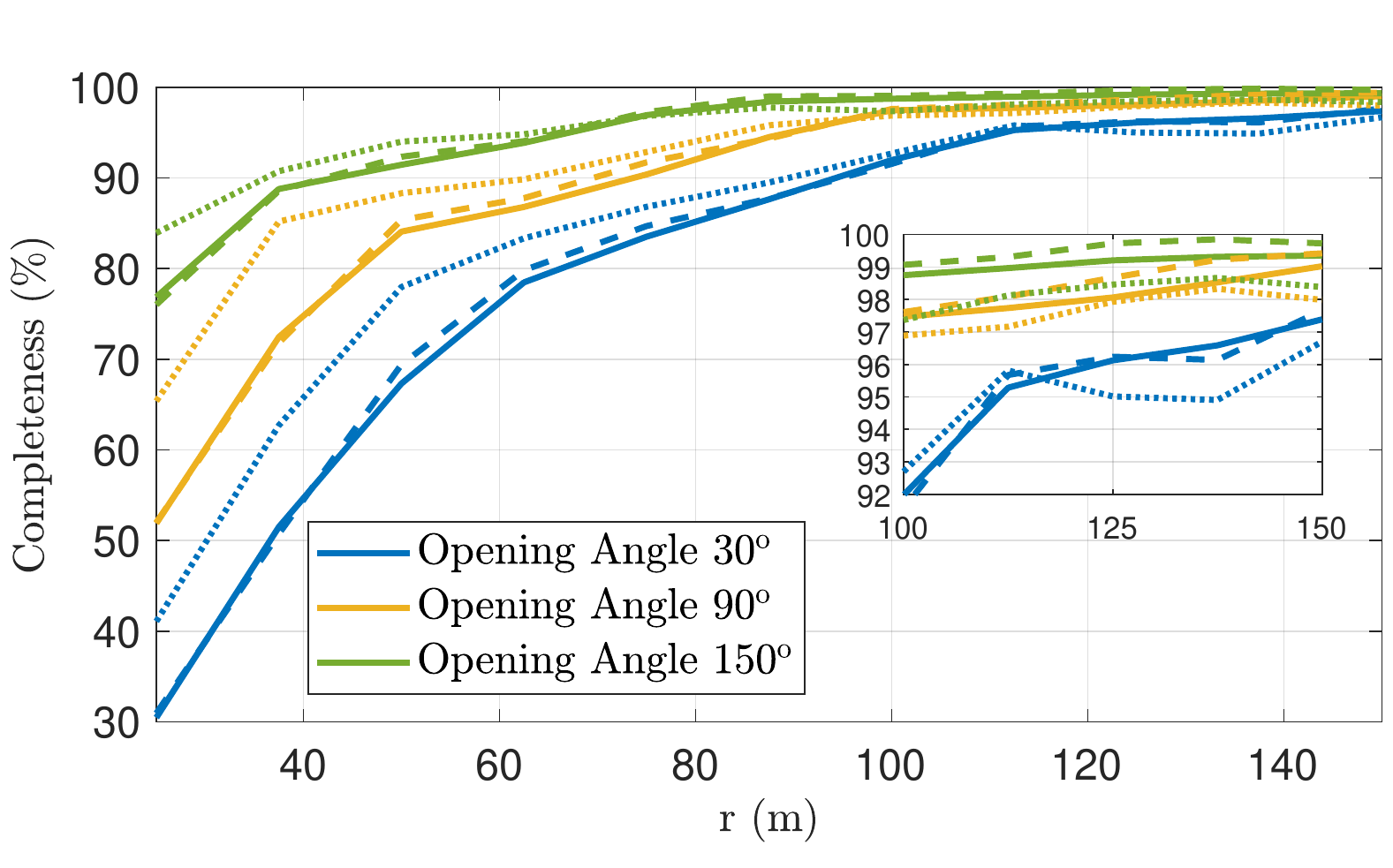}\\
b) Average Tuesday morning\\
\includegraphics[width=1\linewidth]{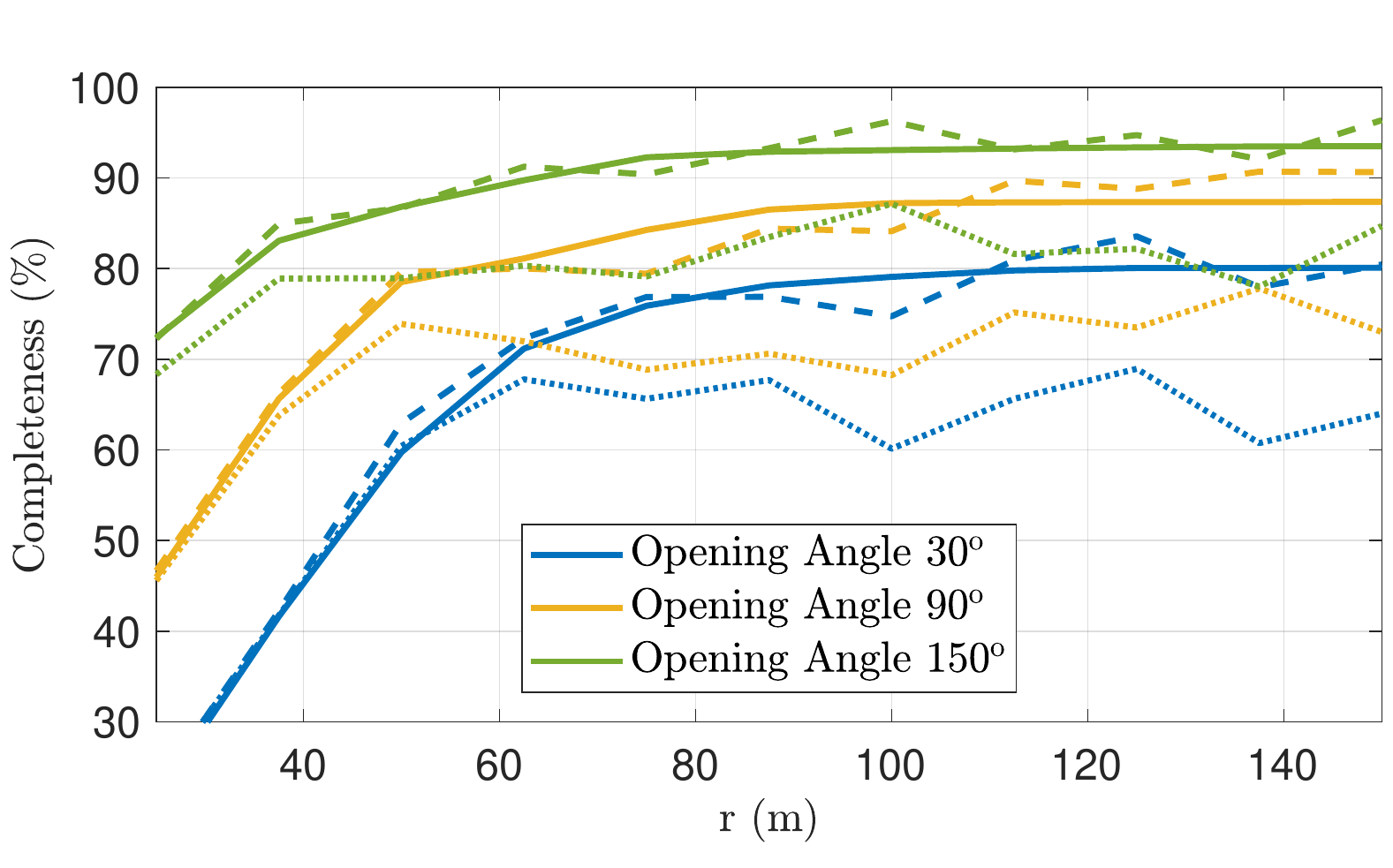}\\
c) Traffic jam
\caption{
Completeness of object detection for selected traffic scenarios, depending on the sensor parameters $r$ and $\omega$.  
We compare perfect camera sensing (full lines), RADAR detections (dashed), and tracking based on RADAR detection (dotted).
For the sake of clarity, only three selected opening angles are presented, yet more have been studied in the course of our analysis.
In scenario $b)$, the emergency lane was open to traffic, while in $a)$ and $c)$ it was closed.
As before we used $\dpyl=50\m$, $\droad=14\m$, $\dsr=0.5\m$, and $\alpha =(\pi -\omega)/2$.
The simulation run-times of each scenarios were adapted to account for a total number of about $100$ passing vehicles.
We chose a sample time of $0.1\s$ for RADAR sensing, and $0.05\s$ for vision sensing.
}
\label{fig:Occlusions}
\end{figure}

Our simulation results are shown in Fig.~\ref{fig:Occlusions}. First, we observe the following general trends.
The degree of completeness of the captured information decreases with increasing traffic density, due to the growing probability of vehicle occlusions.
Furthermore, the perfect cameras and the generic RADAR sensors exhibit very similar detection capabilities, taking into account random fluctuations in the vehicle distribution for a respective traffic scenario. This implies that the postulated difference in resolution does not affect object detection much.
Tracking improves the completeness of detections especially for poor road coverage (e.g. small sensor range), since a vehicle trajectory can still be predicted for some time after the object has left the sensors' FoV. On the other hand, the measurement-to-track association becomes progressively difficult for distant targets, such that detections without tracking perform again better than the KF in the long-range limit.

For a more detailed analysis, we compare the selected traffic scenarios.
In the low traffic density scenario of Fig.~\ref{fig:Occlusions}a (Christmas Eve), with only very few vehicles on the road at the same time, almost all objects are detected by camera, RADAR or RADAR tracking, given that the sensor network provides sufficient road coverage. Detection failures at small sensor ranges are attributed to objects not being in the sensors' FoV. Complete traffic detection up to a degree of $99\%$ can be readily achieved with feasible sensor parameters. 

The scenario in Fig.~\ref{fig:Occlusions}b describes a typical commuting traffic. Here, object occlusions have a noticeable impact on the completeness of the detected information. However, by improving sensor range and opening angle one can eliminate most of the occlusion events, and obtain an average completeness of about $95\%-99\%$. Importantly, for small to moderate sensor ranges, traffic detection benefits greatly from tracking here. 
To clarify this, we note the following mechanism.
A vehicle that is on a distant lane (i.e. a lane that is not closest to the sensor network) experiences alternating periods in which 
there either exists a clear LoS to at least one sensor, or in which it is occluded by another object. Accordingly, a detection of the vehicle is possible or not. 
In this scenario we find that, typically, the intervals in which a selected object is visible are large compared to the track confirmation time, while the occlusion times are small or comparable to the coasting time, such that the filter algorithm can work efficiently.

Last, we study the high-density traffic scenario of a congested highway in Fig.~\ref{fig:Occlusions}c. Due to the increased frequency of occlusions, the degree of completeness achieved in the simulation is reduced, and we find values of about $70\%-95\%$. Note that the KF does not perform very well in this scenario:
In contrast to the preceding cases, the periods in which a selected object is occluded are typically large compared to the KF coasting time, such that the relative benefit of tracking is diminished.
In addition, the increased proximity of vehicles impairs the association of measurements to tracks.
Since the highest vehicle density is usually on the lane which is closest to the sensor network, our system is fairly sensitive to occlusions in this scenario. 
We expect a higher degree of completeness if the sensors are mounted on the opposite, or on either roadside. 
To some extent, the detection capability of the infrastructure can also be improved by refining the tracking parameters.

\section{Summary}
\label{sec:summary}

Assessing the dependability of a roadside infrastructure for AD is a challenge of great complexity, that needs to be tackled from various angles.
Here, we approach this problem by addressing the perceptional completeness of the sensor network in a given setup.
Our analysis offers guidelines for the design of such an infrastructure, e.g. in terms of a favorable sensor positioning.

Since any missing information about an object is potentially safety-critical, to this point no meaningful threshold for the map completeness can be given that was able to guarantee safe automated driving.
Therefore, we here consider the sensor infrastructure as an instrument to support AD, e.g. in the form of augmenting agent-based vision, or for improved traffic orchestration. Those services do not necessarily require an absolute perceptional completeness of the sensor network, yet become more feasible and beneficial with an increasing degree of information completeness.
%
%
We verify the correlation of the latter with the highway traffic density, to demonstrate that the quality of the support the infrastructure can offer is best for low to moderate traffic density scenarios, while in situations of very dense traffic a substantial portion of the information is lost due to vehicle occlusions. 
Note that the sensors were placed only on a single roadside in our model, as such a setup is expected to be of practical relevance. The system design could be easily extended e.g. by deploying sensors on either roadside, or by elevating the sensors, in order to decrease the probability of occlusion events. 
A global design methodology is beyond the scope of this article, however will be a subject of further investigations. The same holds for instance for the analysis of a more refined traffic model, a RCS modeling physical vehicle shapes, or multipath propagation of RADAR waves (e.g. on the road surface for transvision effects).

\section{Acknowledgment}

We thank Frederik Bachmann (Technische Universit\"at M\"unchen) and the Traffic Authority of Southern Bavaria for providing us with traffic measurement data.
This work was funded by the German Federal Ministry of Transport, Building and Urban Development (BMVI) within the project KoRA9 (grant No. 16AVF1032A).

\bibliographystyle{IEEEtran}
\bibliography{FullBib3}

\begin{thebibliography}{10}
\providecommand{\url}[1]{#1}
\csname url@samestyle\endcsname
\providecommand{\newblock}{\relax}
\providecommand{\bibinfo}[2]{#2}
\providecommand{\BIBentrySTDinterwordspacing}{\spaceskip=0pt\relax}
\providecommand{\BIBentryALTinterwordstretchfactor}{4}
\providecommand{\BIBentryALTinterwordspacing}{\spaceskip=\fontdimen2\font plus
\BIBentryALTinterwordstretchfactor\fontdimen3\font minus
  \fontdimen4\font\relax}
\providecommand{\BIBforeignlanguage}[2]{{%
\expandafter\ifx\csname l@#1\endcsname\relax
\typeout{** WARNING: IEEEtran.bst: No hyphenation pattern has been}%
\typeout{** loaded for the language `#1'. Using the pattern for}%
\typeout{** the default language instead.}%
\else
\language=\csname l@#1\endcsname
\fi
#2}}
\providecommand{\BIBdecl}{\relax}
\BIBdecl

\bibitem{Sun2006}
Z.~Sun, G.~Bebis, and R.~Miller, ``{On-Road Vehicle Detection: A Review},''
  \emph{IEEE Trans. on Pattern Analysis and Machine Intelligence}, vol.~28,
  no.~5, pp. 694--711, 2006.

\bibitem{Geiger2012}
A.~Geiger, P.~Lenz, and R.~Urtasun, ``{Are we ready for autonomous driving? The
  KITTI vision benchmark suite},'' \emph{Proceedings of the IEEE Computer
  Society Conference on Computer Vision and Pattern Recognition}, pp.
  3354--3361, 2012.

\bibitem{Kalman1960}
R.~E. Kalman, ``{A New Approach to Linear Filtering and Prediction Problems},''
  \emph{Journal of Basic Engineering, Transactions of the ASME}, vol.~82, no.
  Series D, pp. 35--45, 1960.

\bibitem{Danescu2011a}
R.~Danescu, F.~Oniga, and S.~Nedevschi, ``{Modeling and tracking the driving
  environment with a particle-based occupancy grid},'' \emph{IEEE Transactions
  on Intelligent Transportation Systems}, vol.~12, no.~4, pp. 1331--1342, 2011.

\bibitem{DSRC1999}
``{Federal Communications Commission. FCC 99-305. FCC Report and Order},''
  1999.

\bibitem{Shalev-Shwartz2017}
S.~Shalev-Shwartz, S.~Shammah, and A.~Shashua, ``{On a Formal Model of Safe and
  Scalable Self-driving Cars},'' \emph{arXiv:1708.06374 [cs.RO]}, 2017.

\bibitem{Wu2004}
H.~Wu, R.~Fujimoto, and G.~Riley, ``{Analytical models for information
  propagation in vehicle-to-vehicle networks},'' \emph{2004 IEEE 60th Vehicular
  Technology Conference, 2004. VTC2004-Fall}, vol.~6, no.~C, pp. 4548--4552,
  2004.

\bibitem{Santa2008}
J.~Santa, A.~F. G{\'{o}}mez-Skarmeta, and M.~S{\'{a}}nchez-Artigas,
  ``{Architecture and evaluation of a unified V2V and V2I communication system
  based on cellular networks},'' \emph{Computer Communications}, vol.~31,
  no.~12, pp. 2850--2861, 2008.

\bibitem{Fang2007}
J.~Fang, H.~Meng, H.~Zhang, and X.~Wang, ``{A low-cost vehicle detection and
  classification system based on unmodulated continuous-wave radar},''
  \emph{IEEE Conference on Intelligent Transportation Systems, Proceedings,
  ITSC}, pp. 715--720, 2007.

\bibitem{Felguera-Martin2012}
D.~Felguera-Martin, J.-T. Gonzalez-Partida, P.~Almorox-Gonzalez, and
  M.~Burgos-Garcia, ``{Vehicular Traffic Surveillance and Road Lane Detection
  Using Radar Interferometry},'' \emph{IEEE Transactions on Vehicular
  Technology}, vol.~61, no.~3, pp. 959--970, 2012.

\bibitem{Dickmann2016}
J.~Dickmann, J.~Klappstein, M.~Hahn, N.~Appenrodt, H.~L. Bloecher, K.~Werber,
  and A.~Sailer, ``{Automotive radar the key technology for autonomous driving:
  From detection and ranging to environmental understanding},'' \emph{2016 IEEE
  Radar Conference, RadarConf 2016}, 2016.

\bibitem{Lee2002}
Y.~W. Lee, D.~M. Strong, B.~K. Kahn, and R.~Y. Wang, ``{AIMQ: A methodology for
  information quality assessment},'' \emph{Information and Management},
  vol.~40, no.~2, pp. 133--146, 2002.

\bibitem{Wang1996}
R.~Y. Wang and D.~M. Strong, ``{Beyond Accuracy: What Data Quality Means to
  Data Consumers},'' \emph{Journal of Management Information Systems}, vol.~12,
  no.~4, pp. 5--33, 1996.

\bibitem{Wiltschko2004}
T.~Wiltschko, ``{Sichere Information durch infrastrukturgest{\"{u}}tzte
  Fahrerassistenzsysteme zur Steigerung der Verkehrssicherheit an
  Stra{\ss}enknotenpunkten.}'' \emph{Fortschritt-Berichte VDI}, vol.
  Fortschrit, no. Nr.570, p. 189, 2004.

\bibitem{Cheng2010}
X.~Cheng, P.~Liu, Z.~Chen, H.~Wu, and X.~Fan, ``{The Optimal Sensing Coverage
  for Road Surveillance.}'' \emph{Engineering}, vol.~2, no.~4, pp. 318--327,
  2010.

\bibitem{Dash2018}
D.~Dash, ``{Approximation Algorithms for Road Coverage using wireless sensor
  networks for moving objects monitoring},'' \emph{arXiv:1802.07502}, 2018.

\bibitem{Schopplein2013}
E.~S. Sch{\"{o}}pplein, ``{Integration fahrstreifenbezogener
  Kenngr{\"{o}}{\ss}en und seitlicher Detektionsdaten in ein makroskopisches
  Verkehrsflussmodell f{\"{u}}r dreistreifige Richtungsfahrbahnen},'' PhD
  Thesis, Technische Universit{\"{a}}t M{\"{u}}nchen, 2013.

\bibitem{ABDSB2018}
{Traffic Authority of Southern Bavaria (ABDSB)}, ``{Unpublished Raw Data},''
  2018.

\bibitem{Matlab2017}
\BIBentryALTinterwordspacing
{The Mathworks Inc. (Natick Massachusetts United States)}, ``{MATLAB Automated
  Driving Systems Toolbox Release 2017b},'' 2017. [Online]. Available:
  \url{https://de.mathworks.com/products/automated-driving.html}
\BIBentrySTDinterwordspacing

\bibitem{Richards2014}
M.~A. Richards, \emph{{Fundamentals of radar signal processing}}, 2nd~ed.\hskip
  1em plus 0.5em minus 0.4em\relax McGraw-Hill Education, 2014.

\end{thebibliography}

\end{document}